\documentclass{PoS}

\title{Status of the QPACE Project }

\ShortTitle{Status of the QPACE Project }

\author{%\Large
  H.~Baier$^1$,
  H.~Boettiger$^1$,
  M.~Drochner$^2$,
  N.~Eicker$^{2,3}$,
  U.~Fischer$^1$,
  Z.~Fodor$^3$,
  G.~Goldrian$^1$,
  S.~Heybrock$^4$,
  D.~Hierl$^4$,
  T.~Huth$^1$,
  B.~Krill$^1$,
  J.~Lauritsen$^1$,
  T.~Lippert$^{2,3}$,
  T.~Maurer$^4$,
  J.~McFadden$^1$,
  N.~Meyer$^4$,
  \speaker{A.~Nobile}$^{5,6}$,
  I.~Ouda$^7$,
  M.~Pivanti$^{4,6}$,
  D.~Pleiter$^8$,
  A.~Sch\"afer$^4$,
  H.~Schick$^1$,
  F.~Schifano$^9$,
  H.~Simma$^{10,8}$,
  S.~Solbrig$^4$,
  T.~Streuer$^4$,
  K.-H. Sulanke$^8$,
  R.~Tripiccione$^9$,
  T.~Wettig$^4$,
  F.~Winter$^8$ \\
  $^1$IBM B\"oblingen, $^2$FZ J\"ulich, $^3$Univ. Wuppertal, $^4$Univ. Regensburg, $^5$ECT$^*$ Trento, $^6$INFN Trento, $^7$IBM Rochester, $^8$DESY Zeuthen, $^9$ Univ. Ferrara, $^{10}$ Univ. Milano Bicocca.

}

\abstract{We give an overview of the QPACE project, which is pursuing the
  development of a massively parallel, scalable supercomputer for LQCD. 
  The machine is a three-dimensional torus of identical processing nodes, based
  on the PowerXCell 8i processor.  The nodes are connected
  by an FPGA-based, application-optimized network processor attached
  to the PowerXCell 8i processor.  We present a performance analysis of
  lattice QCD codes on QPACE and corresponding hardware benchmarks.
  }

\FullConference{The XXVI International Symposium on Lattice Field Theory \\
		 July 14 - 19, 2008\\
		 Williamsburg, Virginia, USA}

\begin{document}

\section{Introduction}

The QPACE project is designing and building a novel cost-efficient computer that
is optimized for LQCD applications.  This research area has a long tradition of
developing such computers (see, e.g., \cite{qcdoc,apenext}). Previous
projects  were based on system-on-chip designs, but due to the rising costs of
custom ASICs the QPACE project follows a different strategy: a powerful
commercial multi-core processor is tightly coupled to a custom-designed network
processor. The latter is implemented using a Field Programmable Gate
Array (FPGA), which has several distinct advantages over a custom ASIC: shorter
development time and cost, lower risk, and the possibility to modify the
design of the network processor even after the machine has been
deployed.

The development of QPACE is a common effort of several academic institutions
together with the IBM research and development lab in B\"oblingen (Germany).
First prototype hardware is
already available and currently being tested.

\begin{figure}
\begin{center}
\includegraphics*[scale=0.3,angle=0]{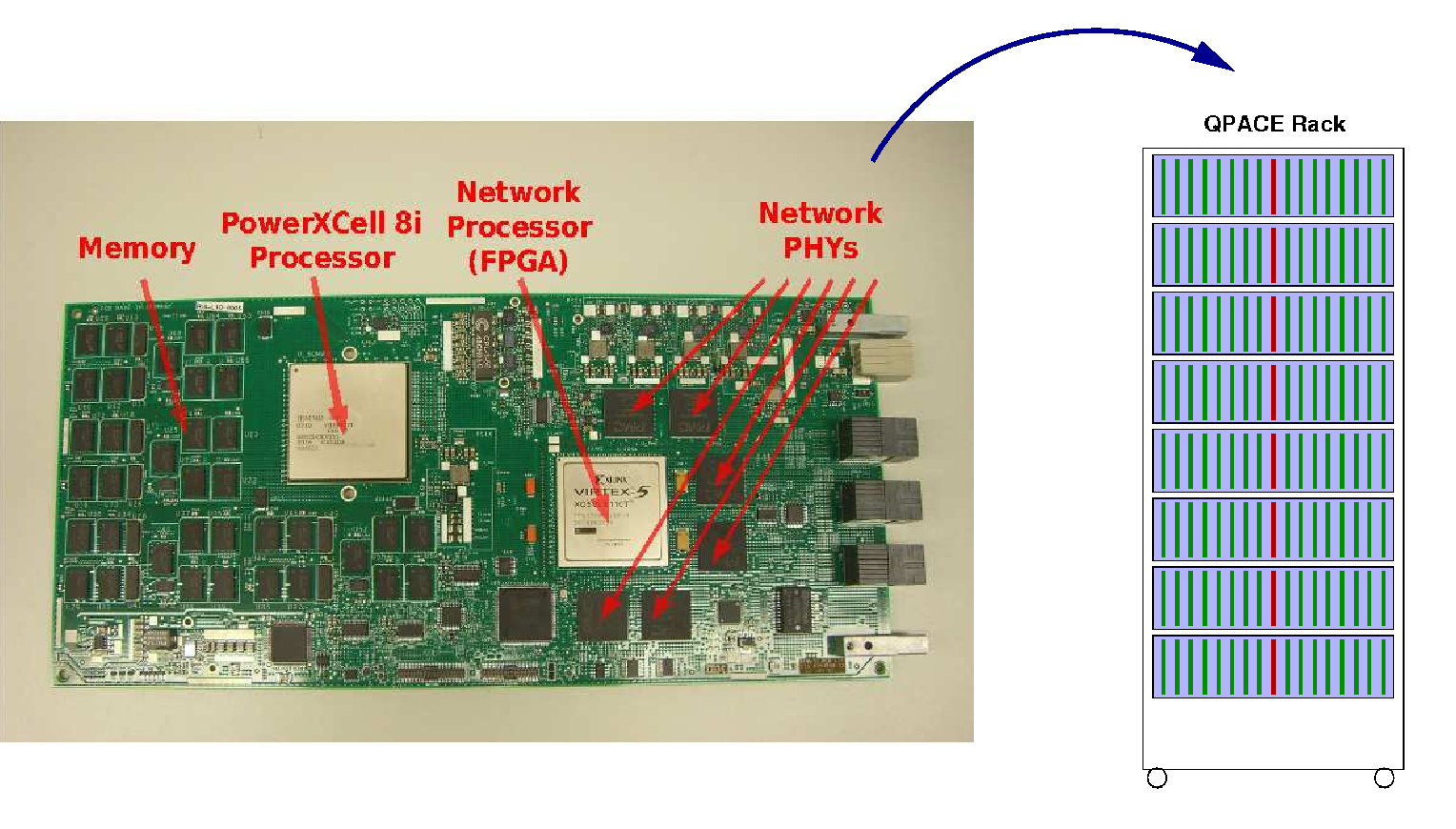}
\caption{The left picture shows a QPACE node card with 4 GBytes of memory,
a PowerXCell 8i processor, an FPGA and 6 torus network PHYs.
256 node cards and 16 root cards are mounted from the front- and rear-side of
a rack (right figure).}
\label{fig_node}
\end{center}
\end{figure}

\begin{figure}
\begin{center}
\includegraphics*[scale=0.3,angle=0]{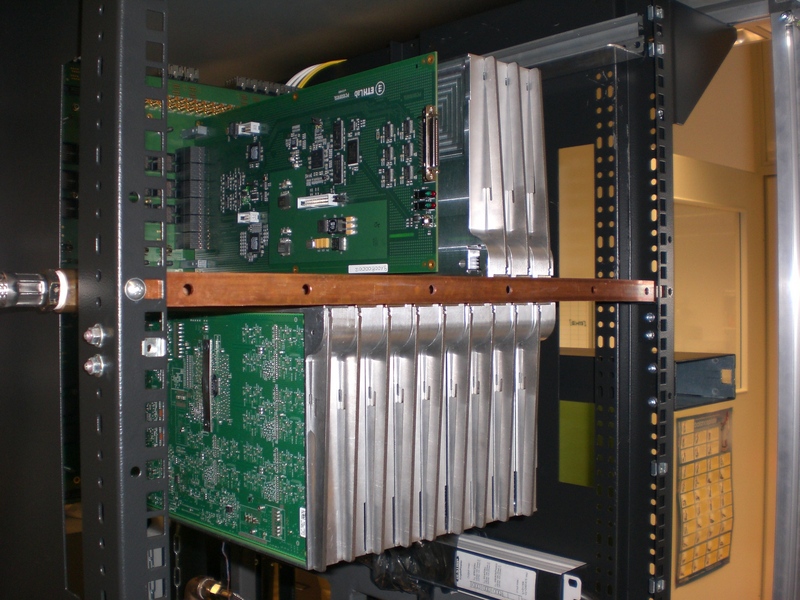}
\caption{Prototype of the QPACE rack with 12 node cards and 1 root card
attached to a backplane. The node cards are assembled in thermal boxes
which conduct the heat to a liquid-cooled plate. A protoype of this
cold plate made of copper is visible in this figure.}
\label{fig_rack}
\end{center}
\end{figure}

\section{The QPACE architecture}

The building block of QPACE is a node card based on IBM's PowerXCell 8i
processor and a Xilinx Virtex-5 FPGA (see Fig. \ref{fig_node}, \ref{fig_rack}). The PowerXCell
8i is the second implementation of the Cell Broadband Engine Architecture
\cite{Hofstee}, very similar to the Cell processor used in the PlayStation 3. The
main reason for using this enhanced Cell processor is its support for
high-performance double precision operations with IEEE-compliant rounding. The
Cell processor contains one PowerPC Processor Element (PPE) and 8 Synergistic
Processor Elements (SPE). Each of the SPEs runs a single thread and has its own
256 kBytes on-chip memory (local store, LS) which is accessed by DMA
or by local load/store operations from/to 128 general-purpose
128-bit registers. An SPE in the PowerXCell 8i processor executes two
instructions per cycle, performing up to 8 single precision (SP) or 4 double
precision (DP) floating point (FP) operations. Thus, the total SP (DP) peak
performance of all 8 SPEs on a single processor is 204.8 (102.4) GFlops
at the nominal clock frequency of 3.2 GHz.

\begin{figure}[ht]
\begin{center}
\includegraphics*[scale=0.40,angle=0]{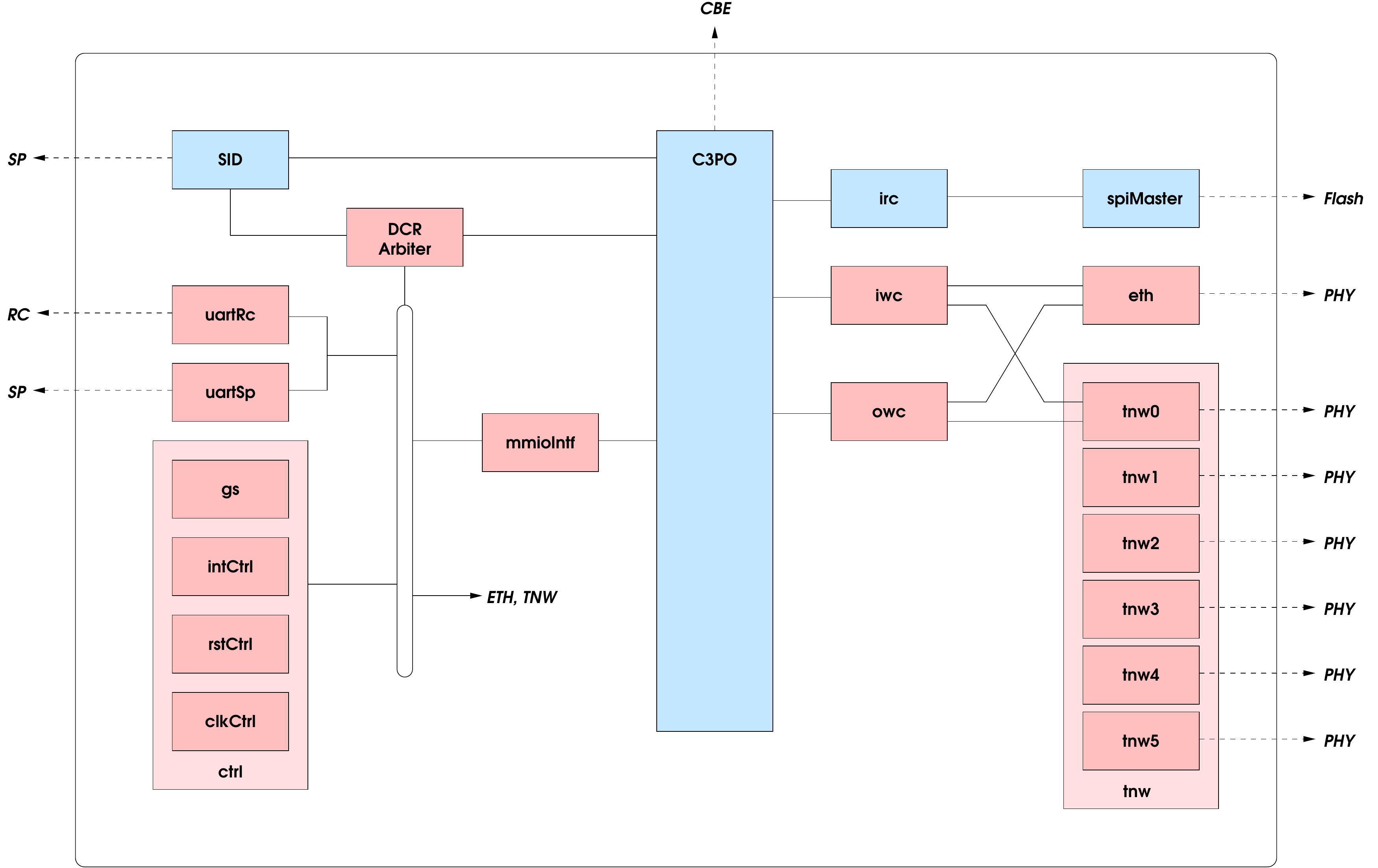}
\caption{Block diagram of the network processor. The C3PO block realizes
the interface to the PowerXCell 8i processor. It provides an IBM proprietary
high-speed bus interface to which the network PHYs are connected (right part).
Most other devices, e.g., ~UART interface to service processor (SP) and root card
(RC) and control blocks, are attached
to a DCR (Device Control Register) bus (left part).}
\label{fig_nwp}
\end{center}
\end{figure}

The processor has an on-chip memory controller supporting a memory bandwidth of
25.6 GBytes/s and a configurable I/O interface (Rambus FlexIO) supporting a coherent
as well as a non-coherent protocol with a total bidirectional bandwidth of up to
25.6 GBytes/s. Internally, all units of the processor are connected to the coherent
Element Interconnect Bus (EIB) by DMA controllers. In QPACE the I/O interface is
used to interconnect the PowerXCell 8i processor with the network processor
implemented on a Xilinx V5-LX110T FPGA (Fig. \ref{fig_nwp}), using 
RocketIO transceivers that are able to support the FlexIO physical layer. 
We plan to use 2 FlexIO links between the multi-core compute processor and the
network processor, with an aggregate bandwidth of 6 GBytes/s
simultaneously in either
direction.

\begin{figure}
\begin{center}
 \includegraphics*[scale=0.6]{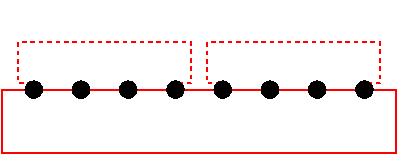}
\caption{Example for the use of redundant network links to change the topology
of the system (here from a single set of 8 to 2 sets of 4 node cards).}
\label{fig_recabling}
\end{center}
\end{figure}

\begin{figure}
\begin{center}
  \includegraphics*[scale=0.4,angle=0]{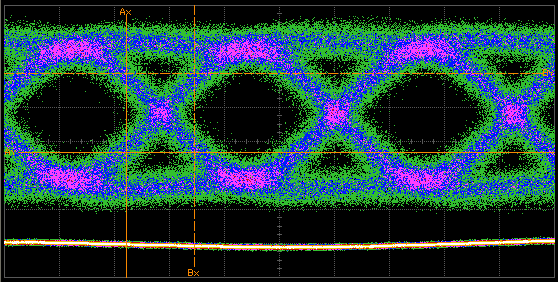}
\caption{A typical eye diagram corresponding to the worst-case
  situation in QPACE, with a 3.125~GHz signal travelling over 50~cm of
  board material, 50~cm of cable, and 4 board connectors.}
\label{fig_eye}
\end{center}
\end{figure}

The node cards are interconnected by a three-dimensional torus network with nearest-neighbor
communication links. The physical layer of the communication links relies on commercial
standards for which well-tested and cheap transceiver hardware components are available.
This allows us to move the most timing-critical logics out of the FPGA.
Specifically, we use the 10 Gbit/s PMC Sierra PM8358 XAUI transceiver. On
top of this standard physical layer we have designed a lean custom protocol
optimized for low latencies. Unlike in other existing Cell-based parallel machines,
in QPACE we can move data  directly from the local store (LS) of any SPE on one node
to the LS of any SPE of one of the 6 neighboring nodes. Data do not have to be
routed through main memory (to avoid the bottleneck of the memory interface) or the PPE.
Rather, data are moved via the EIB directly to or from the I/O interface. The
tentative goal is to keep the latency for LS-to-LS copy operations on the order
of 1$\mu$s. 32 node cards are mounted on a single backplane. One dimension of
the three-dimensional torus network is completely routed within the backplane.
We arrange nodes as one partition of $1\times4\times8$ nodes or as multiple smaller
partitions. For larger partitions, several backplanes can be interconnected by
cables. 8 backplanes sit inside one rack, hosting a total of 256 node cards
corresponding to an aggregate peak performance of 26 TFlops (DP). A system
with $n$ racks can be operated as a single partition with 2$n$ x 16 x 8
nodes.The PMC Sierra PM8358 transceivers have a redundant link interface that
we use to repartition the machine (in two of the three dimensions) without the need for recabling. 
An example of this feature is shown in Fig.
\ref{fig_recabling}. The properties of the physical layer of the network have
been investigated in detail. Figure~\ref{fig_eye} 
displays an example of an eye diagram corresponding to the worst-case
situation in a QPACE machine.

On each node card the network processor is also connected to a Gbit-Ethernet
transceiver. The Ethernet ports of all node cards connect to standard
Ethernet switches housed within the QPACE rack. Depending on the I/O
requirements the Ethernet bandwidth between a QPACE rack and a front-end system
can be adjusted by changing the bandwidth of the uplinks of the switches.

On each backplane there are 2 root cards which manage and control 16 node cards
each (e.g., when booting the machine). The root card hosts a small Freescale
MCF5271 Microprocessor operated using uClinux. The microprocessor can be
accessed via Ethernet, and from it one can connect to various devices on the
node cards via serial links (e.g., UART). The root cards are also part of a
global signal tree network. This network sends signals and/or interrupts 
from any of the node cards to the top of the tree. The combined result  result is then propagated to all node cards of a given partition. 

Each
node card consumes up to 130 Watts. To remove the generated heat a
cost-efficient liquid cooling system has been developed, which enables
us to reach 
high packaging densities. The maximum power consumption of one QPACE rack is about
35 kWatts. This translates into a power efficiency of about 1.5 Watts/GFlops
(DP, peak). 

%\begin{center}
%\begin{table}
%\begin{tabular}{|l|l|l|l|}
%\hline
%nodes & topology  & nodes & topology\\
%\hline
%8	&  $2\times 2 \times 2 $ &256 &  $4\times 8 \times 8 $\\
%32 &  $2\times 4 \times 4 $ & 512 &  $8\times 8 \times 8 $\\
%64 &  $4\times 4 \times 4 $ & 1024 &  $8\times 16 \times 8 $\\
%128 &  $4\times 4 \times 8 $ & 2048 &  $16\times 16 \times 8 $\\
%\hline
%\end{tabular}
%\caption{Is this table really needed: Should we remove it?}
%\end{table}
%\end{center}

\section{Application software and performance}

During an early phase of this project a performance analysis has been
done \cite{lat2007} based on models \cite{Bilardi}
which typically take into account only bandwidth and throughput parameters of the hardware.
The overall performance of LQCD applications strongly depends on how efficiently a few computational tasks can be implemented, namely the application of the lattice Dirac operator on a quark field, and various linear algebra operations. For the hopping term of the Wilson-Dirac operator we
estimated for realistic hardware parameters a theoretical efficiency of about 30\%. The main restrictions come
from the performance of the memory interface. 
While other implementations of the Wilson-Dirac operator code on the
Cell processor exist \cite{Nakamura,epcc}, ours is different.  We worked out a sophisticated strategy for reading data from memory and storing results back to memory, such that external
memory accesses are minimized and a network latency of some $\mu$s,
i.e., ~O(10,000) clock cycles (at a clock frequency of 3.2 GHz),
can be tolerated.

In a real implementation of this application kernel we have
verified that on a single processor an efficiency of 25\% can be achieved with realistic local lattice sizes for the final machine \cite{andrea} (see Table \ref{tab_dirac}).
The Wilson-Dirac operator and the linear algebra routines (see Tables
\ref{tab_la1}, \ref{tab_la2}) have been written using SPE intrinsics and sophisticated addressing techniques.

An efficient implementation of applications on the Cell processor is obviously more difficult compared to
 standard processors. One has to carefully choose the data layout and scheduling of operations in order to optimize utilization of the memory interface. Moreover, the overall performance of the program critically depends on how on-chip parallelization is implemented. To relieve the programmer from the burden of porting efforts we
apply two strategies. For a number of kernel routines which are particularly performance-relevant we
will provide highly optimized implementations which can be accessed through library calls. To facilitate
the implementation of the remaining parts of the code we plan to port or implement software layers
that hide the hardware details.
The user-level API to manage SPE-to-SPE communications via the torus network will be easy to use with complementary send/receive primitives.

\begin{table}
\begin{center}
\begin{tabular}{llcccc}
Lattice   & $L_4\cdot8^3$   &  $L_4\cdot10^3$  \\
   \hline
$L_4=32$  & {{25\%}}   & {{ 24\%}}  \\ 
 $L_4=64$  & {{26\%}}   & {{ 25\%}}  \\ 
$ L_4=128$  & {{26\%}}   & {{ 26\%}}  \\
      \hline
   model   &   34\%   &  34\% \\
\end{tabular}
\caption{Results of the Wilson-Dirac operator benchmark. $L_4$ is the extension of the lattice in which the computation in our scheme proceeds.  See
  Ref.~\cite{lat2007} for details.}
\label{tab_dirac}
\end{center}
\end{table}

\begin{table}
\begin{center}
\begin{tabular}{llcccc}
   version &  caxpy   &  cdot   &   rdot  &   norm \footnotemark[1]\\
   \hline
   
   %xlc/asm     &   {\color{green}{\bf47\%}}   & {\color{green}{\bf42\%}}   &  {\color{green}{\bf38\%}}   & {\color{green}{\bf49\%}}  \\
   xlc/asm     &  47\%  & 42\%  & 38\%   & 49\% \\
   \hline
   model   &   50\%   &  50\%   &   50\%  &   100\%  \\
\end{tabular}
\caption{Linear algebra benchmarks, with data in Local Store.}
\label{tab_la1}
\end{center}
%\end{table}
\end{table}

\begin{table}
%\item Single Precision Benchmarks (main memory)
\begin{center}
\begin{tabular}{llcccc}
version   & caxpy   &  cdot   &   rdot  &   norm \\
   \hline
    xlc  &  3.2\%   &  5.1\%    &  2.7\%   &   5.3\%  \\
% xlc  & {\color{red}{\bf 3.2\%}}   & {\color{red}{\bf 5.1\%}}    & {\color{red}{\bf 2.7\%}}   &  {\color{red}{\bf 5.3\%}}  \\
      \hline
   model   &   4.1\%   &  6.2\%   &   3.1\%  &   6.2\%  \\
\end{tabular}
\caption{Linear algebra benchmarks, with data in Main Memory.}
\label{tab_la2}
\end{center}
\end{table}

\footnotetext[1]{The difference between benchmark and model is due to limited loop unrolling and sub-optimal instruction scheduling.}

\section{Conclusion}

QPACE is a next-generation massively parallel computer optimized for LQCD
applications. It leverages the power of modern multi-core processors by tightly
coupling them within a custom high-bandwidth, low-latency network. The system is
not only optimized with respect to procurement costs vs.~performance but also in
terms of power consumption, i.e., operating costs.

As mentioned above, first prototype hardware is already available.
The test of the final hardware configuration is expected to be
completed by the end of 2008. In early 2009 we plan to start
manufacturing several large machines with an aggregate peak
performance of 200 TFlops (DP). The ambitious goal of the project is
to make these machines available for research in lattice QCD by 
mid 2009.

\section*{Acknowledgments}

QPACE is funded by the Deutsche Forschungsgemeinschaft (DFG) in the
framework of SFB/ TR-55 and by IBM. We thank the developers at the IBM lab in
La Gaude involved in this project and acknowledge important technical
and financial contributions to the project by Eurotech (Italy),
Kn\"urr (Germany), and Xilinx (USA).

\end{document}